\documentclass[a4paper]{article}
\usepackage[utf8]{inputenc}
\usepackage{natbib}
\usepackage{setspace}
\usepackage{amsmath}
\usepackage{multirow}
\usepackage{lscape}
\usepackage{geometry}
\usepackage{graphicx}
\title{Inferring ancestral states\\ without assuming neutrality or gradualism \\using a stable model of continuous character evolution}
\author{Michael G. Elliot and Arne \O. Mooers \\ Department of Biology\\and Human Evolutionary Studies Program\\Simon Fraser University, 8888 University Drive Burnaby\\BC V5A 1S6, Canada}
\date{}

\begin{document}

\maketitle

\begin{abstract}
The value of a continuous character evolving on a phylogenetic tree is commonly modelled as the location of a particle moving under one-dimensional Brownian motion with constant rate. The Brownian motion model is best suited to characters evolving under neutral drift or tracking an  optimum that drifts neutrally. We present a generalization of the Brownian motion model which relaxes assumptions of neutrality and gradualism by considering increments to evolving characters to be drawn from a heavy-tailed stable distribution (of which the normal distribution is a specialized form). We describe Markov chain Monte Carlo methods for fitting the model to biological data paying special attention to ancestral state reconstruction, and study the performance of the model in comparison with a selection of existing comparative methods, using both simulated data and a database of body mass in 1,679 mammalian species. We discuss hypothesis testing and model selection. The new model is well suited to a stochastic process 
with a volatile rate of change in which biological characters undergo a mixture of neutral drift and occasional evolutionary events of large magnitude.
\end{abstract}

Statistical methods that take into account the dependencies introduced into comparative data by phylogenetic relatedness are fundamental to hypothesis testing and exploration in comparative biology \citep{HarveyPagel91,Martins96}. Each comparative method implicitly imputes to the evolutionary process some specific stochastic model \citep{PagelHarvey89}. The validity of phylogenetic comparative methods depends on the degree to which historical events can be accommodated by the underlying stochastic model of trait evolution, and a mismatch between model and reality can yield erroneous statistical results, especially in the reconstruction of inaccurate ancestral character states \citep{Cunningham1999,OakleyCunningham2000,FinarelliFlynn2006,Slateretal2012}. For this reason it is important to develop 
realistic stochastic models of character evolution (and to constrain those models using empirical data where possible).

The Brownian motion model of evolution -- in which the value of a continuous trait evolves by accruing incremental changes drawn from a random distribution with zero mean and finite constant variance, such that the sum of many increments is distributed according to 
a normal density \citep{HarveyPagel91} -- was introduced to model changes in gene frequencies by \citet{CavalliSforzaEdwards67} but now underlies (directly or indirectly) a range of popular methods for the analysis of continuous traits distributed over phylogenetic trees. In the realm of ancestral state reconstruction the value of a trait evolving across a phylogenetic tree can at all times be shown to be probabilistically distributed according to a normal distribution with variance depending only on tree topology and branch lengths \citep{Maddison91,McArdleRodrigo1994,Schluteretal97}, while in the realm of regression analysis the model yields normally distributed residuals and a covariance matrix 
depending only on tree topology and branch lengths \citep{Grafen89}, both cases giving rise to simple and analytically-tractable solutions. The Brownian walk of a trait value can be regarded as a model of gradualistic neutral evolution since variation in the trait arises from a process of random drift over the branches of a phylogeny at a constant rate and without directionality. A further application thus involves identifying significant departures from the expectation of the Brownian motion model as a means of detecting adaptive variation in the tempo and mode of trait evolution \citep{Harmonetal2003,Harmonetal2010,OMearaetal06}.

We here describe a simple generalization of the Brownian motion model of continuous character evolution which extends the model to include cases where increments to an evolving character may arise from a symmetric stochastic process but without assuming constant finite variance. Not only is the Brownian assumption of constant finite variance, to the best of our knowledge, unverified in existing biological systems, but its relaxation may better suit the domain of application of the standard Brownian 
motion model, namely biological characters likely subject to some degree of selection, and in many cases it may offer a more robust form of statistical inference with respect to outliers in the data. Broadly speaking, we conceptualize diversifying selection on continuous characters as causing an increase and purifying selection on such characters a decrease in the rate of evolutionary change. The relative frequencies of these forms of natural selection along with neutral drift are expected to generate --- for sums 
of 
evolutionary 
increments over long periods of time --- limit distributions with heavier tails than expected under the Brownian motion model. 

Just as the limit distribution of sums of variates drawn from a distribution with constant finite variance is the normal distribution, so the limit distribution of sums of variates drawn from a distribution \textit{without} fixed finite variance is the stable distribution \citep{GnedenkoKolmogorov49, SamorodnitskyTaqqu94, UchaikinZolotarev99}, which has the normal distribution as a special case and which otherwise is characterized by heavy tails, closure under convolution and, potentially, by skew \citep{Nolan97}. Stochastic processes and random walks with volatile variance and heavy tails have previously been modelled robustly using stable distributions in areas as diverse as the study of fractional diffusion \citep{Gorenfloetal02}, earthquake forecasting \citep{Lavallee03}, signal processing \citep{Nikias95}, animal foraging \citep{Viswanathanetal96}, rainfall modelling \citep{MenabdeSivapalan00}, commodity pricing \citep{Weron06}, real estate markets \citep{Young08}, foreign exchange rates \citep{
FofackNolan01}, financial statistics \citep{Rachev03}, image processing \citep{Arce05} and telecommunications management \citep{Mikoschetal02}. \citet{Landisetal2013} recently used a Brownian motion model with L\'{e}vy stable jumps to model jumps in the evolution of continuous traits. According to our view of selection resulting in the generation of evolutionary rate volatility we limit our attention in this paper to the symmetric zero-centred stable distribution parameterized by $\alpha$, the index of stability and $c$, the scale. We model evolution using the stable generalization of Brownian motion, the stable random walk \citep{SamorodnitskyTaqqu94}. The stable random walk shares some attractive properties of Brownian motion, the most important being closure under convolution, such that the sum of several stable distributions is itself a stable distribution with the same $\alpha$ parameter. This closure under linear transformation contrasts with all other non-stable heavy tailed 
distributions, which may yield complex and analytically intractable mixtures in convolution. Thus, after accumulating increments from a symmetrical zero-centred stable distribution, the value of an evolving trait is always probabilistically distributed according to a stable distribution with mean equal to the trait value prior to accumulation and with scale proportional to the number of increments, or in phylognetic parlance the branch length.

Figure 1 illustrates the log probability densities of some unit-scale zero-centred symmetrical stable distributions that differ in the value of $\alpha$, including the special cases of the normal distribution ($\alpha=2$) and the Cauchy distribution ($\alpha=1$).  Figure 2 illustrates some stable random walks driven by accumulation of increments from stable noise with various $\alpha$ values. It is clear from inspection of these figures that declining $\alpha$ is associated with increasingly heavy tails, translating into increasingly volatile random walks which exhibit Brownian-like drift (associated with increments drawn from the high-probability modal region of the underlying distribution) interspersed with occasional rapid jumps in trait value (associated with increments drawn from the low-probability heavy tails). 

As a motivating example of how the accommodation of evolutionary rate volatility may affect the inference of evolutionary processes, we present a simple toy model of continuous character evolution on a phylogenetic tree with six tips. The values of an evolving character were simulated under the Brownian motion model and values at internal nodes were reconstructed based on values at the tips only, first by fitting a Brownian motion model and second by fitting a stable model using the method presented later in this paper. Next, the value of the character on a single tip was artificially inflated by a factor of ten, to represent a bout of adaptive evolution (or other event such as measurement error) on the branch leading to that tip. Again, values at internal nodes were reconstructed under both models.  Results are illustrated in Figure 3. Both reconstruction methods yield similar ancestral states for the 
original Brownian motion data, but differ for the manipulated data. The Brownian motion model exhibits an 
``averaging effect'' in which the 
apparently high rate of evolution resulting from the manipulation is distributed somewhat evenly over all the 
branches, causing a large increase in estimated ancestral states for several internal nodes. The stable model, however, can entertain rare increments of large magnitude, and so is not strongly affected by the manipulation; the apparently high rate of evolution on a single branch can be accommodated as a rare event in a heavy-tailed stochastic process.

In this paper we describe the stable model and the procedures used to fit it to empirical data, with an emphasis on ancestral state reconstruction. We outline strategies for hypothesis testing and model selection. We apply the stable model to simulated data in order to estimate error rates of the model selection procedure, and also use it, alongside a number of alternative models, in an illustrative case study of ancestral state reconstruction based on a a large dataset of mammalian body masses. Finally we discuss the relationship between the stable model of trait evolution and a number of alternative extensions of the Brownian motion model that have been proposed in the literature, and suggest avenues for further development.

\section*{Methods}

\subsection*{Model}

Consider a rooted phylogenetic tree $\mathcal{T}$ which may or may not contain polytomies. A continuous
character \textbf{X} evolves along the branches of $\mathcal{T}$, taking values $b_1$ and $b_2$ at the beginning and end, respectively, of each branch $b$. Under the standard Brownian motion model of evolution, the continuous character evolves by accumulating random independent increments drawn from a probability distribution with constant mean zero and constant finite variance $\sigma^2$. According to the central limit theorem, the sum of such increments along a branch $b$ of length $t_b$ is probabilistically distributed according to a normal density with mean zero and variance $t_b\sigma^2$, a density which we denote $\phi(b_2 - b_1; t_b \sigma^2)$. Given the independence of increments, and therefore of branches, the likelihood of an ancestral state reconstruction of a continuous character evolving under Brownian motion is given by the product:
\begin{equation} \label{eq:BrownianL}
\mbox{L}(\textbf{X}, \sigma; \mathcal{T}) = \prod_{b} \phi (b_2-b_1; t_n\sigma^2)
\end{equation}

If the variance of the increment generating distribution is not constant and finite (as we suppose to be the case under departures from neutrality and gradualism) then according to the generalized central limit theorem the limit distribution for the sum of random independent variates is not normal but falls into the more general class of stable distributions, parameterized by an index of stability $\alpha$ and a scale $c$. The symmetric stable distribution has probability density denoted $\mbox{S}(x;\alpha,c)$. Following \citet{MatsuiTakemura04}, the unit stable density with $c=1$ may be defined as:
\begin{equation} \label{eq:StablePDF}
\mbox{S}(x;\alpha,1)=\dfrac{\alpha}{\pi |\alpha-1| x}\int_0^{\frac{\pi}{2}} \mbox{G}(\kappa; \alpha,x) \exp (-\mbox{G}(\kappa; \alpha,x) ) d\kappa
\end{equation}
where
\begin{equation}
\mbox{G}(\kappa;\alpha,x)=\left(\dfrac{x \cos \kappa}{\sin \alpha \kappa}\right)^{\frac{\alpha}{\alpha-1}} \dfrac{\cos(\alpha-1)\kappa}{\cos \kappa},
\end{equation}
and the general symmetrical stable density $\mbox{S}(x;\alpha,c)$ is a transformation of the unit stable density $\mbox{S}\left(\frac{x}{c};\alpha, 1\right)/c$.

One important property of the stable distribution is that the normal distribution is a special case with $\alpha=2$. For the zero-centred symmetrical cases treated here, we note that:
\begin{equation}\label{eq:normalSpecialCase}
\phi(x;\sigma^2) = \mbox{S}\left(x;2,\frac{\sigma}{\sqrt{2}}\right)
\end{equation}
Furthermore, the sum of $t$ variates drawn from a stable distribution $\mbox{S}(\alpha,c)$ is distributed as $\mbox{S}\left(\alpha, (t c^\alpha)^\frac{1}{\alpha}\right)$. Thus, under a stable model of evolution, the likelihood of an ancestral state reconstruction of $\textbf{X}$ is given by:
\begin{equation}\label{eq:StableL}
\mbox{L}(\textbf{X}, \alpha,c;\mathcal{T}) = \prod_{b} \mbox{S}\left(b_2 - b_1; \alpha, (t_b c^\alpha)^\frac{1}{\alpha}\right)
\end{equation}
which is functionally identical to Equation \eqref{eq:BrownianL} when $\alpha=2$.

Unfortunately, there is no analytical solution to the stable probability density function in Equation \eqref{eq:StablePDF}, so it is necessary to employ numerical methods \citep{BrorsenYang90, Nolan97, McCulloch98, Nolan01, MatsuiTakemura04} to calculate likelihoods of stable models. The model does not lend itself to direct maximum likelihood estimation of parameters due to the existence of a highly multimodal likelihood surface and because arbitarily high likelihoods can be obtained by setting $b1=b2$ for any single internal branch $b$ and having $c\rightarrow0$, a problem exhibited by other statistical models with non-constant variance \citep{Ciupercina2003}, and circumvented through the placing of an appropriate prior on the scale parameter that penalizes the approach to zero, and fitting the model using a Bayesian approach. Since numerical estimation methods are unreliable under extremely heavy tails (i.e. $\alpha<0.2$) \citep{MatsuiTakemura04} we apply flat or triangular priors to the index of stability 
on the domain $0.2<\alpha\leq2$, and a loose inverse gamma prior on 
the scale parameter which has $Pr(c\rightarrow0)\rightarrow0$.

\subsection*{MCMC estimation}

Markov chain Monte Carlo (MCMC) methods are widely used to estimate complex multivariate probability densities in numerous biological fields. The goal of such methods is to generate a sample from a probability distribution by constructing a Markov chain that has the desired distribution as its equilibrium density. A common strategy is to utilize a Metropolis-Hastings sampler \citep{Metropolis53} in which the statistical model is initialized with some set of parameter values $\theta$, a new 
candidate parameter $\theta^\prime$ is generated by a 
symmetrical proposal distribution, and 
accepted as the next step of the Markov chain with probability equal to $\mbox{P}(\theta^\prime)/\mbox{P}(\theta)$. We found that the Metropolis-Hastings sampler performed poorly in estimating ancestral states and parameters of the stable model due to the multimodality of the local likelihood surface (Figure 4) and the non-independence of ancestral state values, which together generate numerous very small ``islands'' of high likelihood which are unlikely to be explored by the Markov chain in a reasonable amount of time. Modified versions of the procedure, such as Metropolis-coupled Markov chain Monte Carlo \citep{Geyer91} did not yield any benefit. 

We found that implementation of a slice sampler \citep{Neal03} manifestly improved the mixing of Markov chains. At each step of the chain, the value of each individual parameter $\theta_i$ is replaced by a new value $\theta_i^\prime$ drawn randomly from the conditional probability distribution $\mbox{P}(\theta_i^\prime | \theta_j, \theta_k, ...)$. This is accomplished in two steps. First, the conditional probability of $\theta_i$ is calculated and a random number $y$ is generated from the uniform distribution between zero and $\mbox{P}(\theta_i | \theta_j, \theta_k, ...) $. Second, $\theta_i^\prime$ is generated as a uniformly distributed random number from the set of ranges for which $\mbox{P}(\theta_i | \theta_j, \theta_k ...) > y$. Even though the conditional probability distribution of the ancestral state of some focal node may be multimodal (Figure 4), the number of modes is known to be less than or equal to the number of nodes to which the focal node is connected by branches.  In addition, 
the set of 
ranges of values for which $\mbox{P}(\theta_i | \theta_j, \theta_k, ...) > y$ can be found by searching around the values of those modes, which are near to the values of the current ancestral states estimated for the nodes to which the focal node is connected. Critically, this sampling procedure (Figure 5) permits large jumps away from suboptimal likelihood peaks even when the sampled distribution is multimodal with widely separated modes. 

\subsection*{Hypothesis testing and model selection}

It is desirable to formulate a statistical model selection criterion to determine whether the stable model of continuous character evolution (with $\alpha<2$) fits the data better than than the Brownian motion model (with $\alpha=2$), as a means of estimating the best possible ancestral state estimates and of testing the hypothesis that a set of character data at the tips of a phylogeny exhibits patterns consistent with departure of the evolutionary process from neutrality. Our Markov chain Monte Carlo procedure generates a sample from the posterior distribution of stable parameters and ancestral states, from which it is difficult to calculate Akaike's Information Criterion (AIC) \citep{Akaike74} due to its reliance on maximized likelihood. However there exists a number of Bayesian generalizations of AIC which may be calculated from posterior MCMC samples, notably the Deviance Information Criterion (DIC) \citep{Spiegelhalteretal2002} which is constructed from the deviance of posterior samples and the 
effective number of parameters in the model. For each individual sample in the MCMC chain the deviance $\mbox{D}(\textbf{X}, \alpha,c;\mathcal{T})$ is equal to $-2 \mbox{ log } \mbox{L}(\textbf{X}, \alpha,c;\mathcal{T})$ plus some unknown normalizing constant which cancels out in comparisons across a pair of models. The effective number of parameters $p_D$ is measured across the whole set of samples and is defined as $\bar{\mbox{D}}-\hat{\mbox{D}}$, where $\bar{\mbox{D}}$ is the mean deviance and $\hat{\mbox{D}}$ is the deviance of the mean parameter values. The Deviance Information Criterion itself is defined as $\mbox{DIC} = \bar{\mbox{D}} + p_D = \hat{\mbox{D}} + 2 p_D$, with lower DIC values indicating better model fit. The  
effective number of parameters $p_D$ may be intuitively understood as a measure of dispersion of parameter estimates around their mean values with respect to likelihood, which increases with model complexity and which acts as a penalty on the mean likelihood of the posterior sample as a whole. Our simulation studies indicated that in phylogenetic datasets $p_D$ did not increase sufficiently in line with tree size, resulting in overfitting of stable models on large trees. A Bayesian Predictive Information Criterion (BPIC) developed by Ando \citep[][personal communication]{AndoTsay2010}, which amounts to DIC with a increased multiplier on the $p_D$ penalty, resolved this problem of overfitting. The simulation studies are available on the software website (see below).

\subsection*{Software implementation}

Efficient software for fitting the stable model to phylogenetic trees and their associated data was written in C++ and is available at http://www.sfu.ca/\textasciitilde micke/stabletraits.html as source code and also compiled for various operating systems. The sofware reports a posterior sample of ancestral state reconstructions and stable parameter values in a format compatible with the Tracer software application \citep{Tracer}, along with the proportional scale reduction factor convergence diagnostic \citep{BrooksGelman1998} and Bayesian Predictive Information Criterion for assessment of model fit \citep{AndoTsay2010}. Multiple chains are run on independent threads, or on independent processors in a cluster computing environment (for which a torque job submission script is also available). 

\subsection*{Application to simulated and natural data}

In order to assess the improvement (if any) in quality of ancestral state reconstruction and the ability of statistical tests to identify biological characters that have evolved under a stable rather than Brownian stochastic process, data were simulated under a variety of conditions. Evolutionary increments were generated randomly from a stable model with unit scale and index of stability ranging from 1.0 to 2.0 in steps of 0.2, on random phylogenetic trees with 25, 40, 60, 90, 130, 175, 235, 325, 440 and 600 tips generated under the Yule 
model in Mesquite \citep{Mesquite}. This procedure gave rise to 60 experimental conditions, each consisting of 250 trees and simulated datasets, and varying in tree size and index of stability of simulated trait values. Based on data at the tips, ancestral states were reconstructed using the modal posterior density estimate from MCMC, first with $\alpha$ fixed at 2.0 (representing a Brownian motion model) and second with a free $\alpha$ (representing a general symmetric stable model). Ancestral states were also esimated using the homogenous Ornstein-Uhlenbeck model for comparison. Estimates of Type I and Type II error rates under the BPIC criterion were made for each tree size/stability condition. Accuracy was assessed for Brownian, Stable and OU models by calculating variance of the inferred ancestral states from the true simulated states for each simulated dataset under each condition. We report here the median variance ratio of stable/Brownian and stable/Ornstein-Uhlenbeck 
reconstructions within each experimental condition.

To provide a demonstration of the model's application to real biological datasets, we made use of a supertree of mammalian species \citep{Bininda-Emonds07, Bininda-Emonds08} and fitted the stable model of character evolution to data on the log adult body mass of 1,679 eutherian mammals \citep[combined data from ][]{Ernest03,Pitnick06}. For purposes of comparison we also estimated ancestral states under two homogenous evolutionary models, the Brownian motion model and the Ornstein-Uhlenbeck model, the latter estimates obtained by maximum likelihood search from parameter estimates generated by the SLOUCH package in R \citep{Hansenetal2008}. Finally we also inferred ancestral states using two heterogenous models, the time-heterogenous Early Burst model  \citep{Blomberg2003,Harmonetal2010} in which the rate of a Brownian process increases or decreases exponentially in time, and the clade-heterogenous model of \citet{Eastmanetal2011} in which the rate of a Brownian process is ``inherited'' within clades but 
allowed to 
occasionally shift in value on probabilistically selected branches of a phylogeny. The former ancestral state estimates were made by maximum likelihood search based on parameter estimates and scaled phylogenies derived from the GEIGER package in R \citep{Harmonetal2008} while the latter ancestral state estimates were made by Brownian motion maximum likelihood reconstruction on a tree with branch lengths scaled by the maximum posterior probability estimates of rates reported by the Auteur package in R \citep{Eastmanetal2011}. Connections between these various models and the stable model are discussed below.

\section*{Results}

Our analysis of simulated evolution on random phylogenetic trees indicates that biological traits derived from a Brownian process can be statistically distinguished from those derived from a non-Brownian symmetrical stable process by on the basis of the Bayesian Posterior Information Criterion (Table \ref{BPIC}). This model selection criterion was found to be highly conservative, with a low rate of false rejection of the null hypothesis for trees of all sizes, but with the expected low power to detect small departures from the neutral Brownian model on small trees. 

In ancestral state reconstruction of traits simulated under Brownian motion the stable model performed as well as the Brownian model (Table 2), with median squared error ratio ranging from 1.02 to 1.00, and better than the Ornstein-Uhlenbeck model, with median squared error ratio ranging from 0.36 on the smallest tree to 0.86 on the largest (Table 3). For trees simulated under the stable process with $\alpha<2$, the stable model yielded more accurate ancestral state estimates, increasingly so for large trees and large deviations from Brownian motion. For the most extreme index of stability considered here ($\alpha=1.0$) the mean squared error under Brownian motion reconstruction was from 4.8 to 100 times higher, and under Ornstein-Uhlenbeck reconstruction from 5.9 to 100 times higher, than the mean squared error under stable reconstruction. 

Results of our analysis of a dataset of mammalian body masses are presented in Table 4. The maximum posterior probability estimate of the index of stability $\alpha$ was 1.55 and the Brownian motion model was soundly rejected in favour of the stable model under the BPIC model selection criterion ($\Delta$BPIC=465). The Ornstein-Uhlenbeck  also fit significantly better than the Brownian motion model ($\Delta$AICc=24) as did the multi-rate model of \citet{Eastmanetal2011} (posterior probability of no rate changes = 0). In order to provide the entries in Table 4 for Early Burst we inferred maximum likelihood ancestral states under the global parameter estimate reported by \citet{CooperPurvis2010} in their broader study of mammalian body mass evolution: the Early Burst model did not differ significantly from the Brownian motion model on this dataset. 

\section*{Discussion}

Reconstructing a historical narrative of trait evolution over time is central to both the formulation and testing of hypotheses in evolutionary biology \citep{Coddington1988,Pagel1999,Martins2000}. Comparative phylogenetic methods do so in a formal framework using stochastic models of the evolutionary process that implicity or explicity assume some probabilistic distribution of ancestral states over internal nodes of a phylogeny  \citep{PagelHarvey89,HarveyPagel91,Martins96,Mesquite,Pageletal2004}. Brownian motion is a fundamental stochastic model of evolution which assumes that biological traits evolve by accruing incremental changes drawn from a random distribution with zero mean and finite constant variance. However, most evolutionary hypotheses of interest involve traits thought to be subject to selection leading to directional tendencies, relatively rapid grade shifts and convergent evolution.  The resulting patterns may be at odds with the neutral drift modelled by Brownian motion \citep{Cunningham1999}
. Indeed, studies of the performance of ancestral state reconstruction using known ancestral states derived from fossil estimates \citep{FinarelliFlynn2006,Donoghueetal1989,WebsterPurvis2002,Slateretal2012} or from taxa evolving sufficiently rapidly to be observed in real time \citep{OakleyCunningham2000} indicate that a mismatch between the stochastic model and historical reality can result in incorrect estimates \citep[but see also][]{
Polly2001}.

In this paper we have described a stochastic model of continuous character evolution based on a generalization of the Brownian model of evolution that does not assume that the rate of evolutionary change is constant and finite. Under these relaxed assumptions, the sum of increments accruing to an evolving character along each branch of a phylogeny is known to tend toward a stable limit distribution, which is identical to a normal distribution in the special case of Brownian motion but otherwise has heavier tails (Figure 1). These heavy tails allow rare evolutionary increments of large magnitude to occur, resulting in a volatile evolutionary process characterized by occasional ``adaptive'' evolutionary shifts interspersed with neutral-like patterns of variation (Figure 2). Stable parameters and ancestral states can be fit to biological data distributed over a phylogenetic tree using Markov chain Monte Carlo methods. We have implemented software that makes use of a slice sampler \citep{Neal03} to sample the 
posterior probability distribution of ancestral state assignments at each node and the values of stable parameters (the index of stability $\alpha$, which is equal to 2 under Brownian motion, and scale $c$). The slice sampler is able to take advantage of our knowledge of the approximate location of modal regions to move across a multimodal likelihood surface without becoming trapped in locally but not globally optimal regions (Figure 5). An additional benefit of the slice sampler is its adaptive step size, requiring no tuning of proposal distributions, which makes practical application of the method straightforward. Our analysis of simulated data indicates that the Bayesian Predictive Information Criterion (BPIC) provides a conservative test of the hypothesis of departures from neutrality (i.e., the existence of heavy tails) in an evolutionary process (Table \ref{BPIC}), and that the stable model estimates ancestral states with reduced error in comparison with the Brownian motion model, when traits evolve by 
accumulating increments from a probability distribution without constant finite variance (Table \ref{BrownianError}). 

We found the stable model to fit the eutherian body-size data significantly better than the Brownian motion model (Table 4), suggesting the existence of departures from neutrality. Under the stable model, the ancestral eutherian is relatively small; in line with fossil evidence (also presented in Table 4), low body mass persists through early diversification of the Superorders Afrotheria, Euarchontoglires and Laurasiathera and the origin of various orders of small size such as Primates, Rodentia, Lagomorpha, Scandentia, Afrosoricida and Macroscelidea; large reductions in body size are rare, occuring in Chiroptera, while large increases in body size occur with 
the origin of several modern Orders of relatively large species including the ungulates (Perissodactyla + Artiodactyla), Carnivora, Cetartiodactyla, Proboscidea and Sirenia.  The Brownian motion model differs in several respects. First the Brownian motion reconstruction of the ancestral eutherian's body mass is an order of magnitude greater than the stable reconstruction; the Brownian motion reconstruction thus posits significant reductions of body size prior to the evolution of orders of small body size including Rodentia, Scandentia, Chiroptera, Eulipotyphla and Macroscelidea. As expected, the Brownian motion model exhibits an ``averaging effect''  more generally, in which transformations in body mass are distributed somewhat evenly over the phylogeny, while the stable model permits large transformations in body mass to occur on a smaller subset of branches. For this reason, the ancestral state reconstruction of taxa ancestral to typically large species (i.e., Cetartiodactyla, Perissodactyla, Proboscidea) 
are smaller 
under the Brownian motion model, and the ancestral state reconstruction of taxa ancestral to typically small species (i.e., Rodentia, Lagomorpha, Chiroptera, Eulipotyphla, Afrosoricida, and Macroscelidea) are larger under the Brownian motion model. The tendency for ancestral state reconstructions to be weighted toward intermediate values is stronger under the Brownian motion model than under the stable model since the former model vitiates against the inference of directional tendency.

A desire to model the adaptive evolution of continuous traits has given rise to a number of approaches that refine or extend the Brownian model. We categorize these as homogenous approaches, in which the stochastic process underlying the generation of evolutionary increments to an evolving character does not vary across branches of a phylogenetic tree, versus heterogenous processes, in which the stochastic process varies across branches. One popular homogenous model is the global Ornstein-Uhlenbeck model of trait evolution \citep{Hansen1997}, in which the direction and rate of evolution at any time depends upon a selection coefficient and the degree of deviation of the trait's current value from some global optimum or ``phylogenetic mean''. This so-called ``mean-reverting'' process has been used as a model of stabilizing selection since deviation away from the phylogenetic mean is penalized under maximum likelihood reconstruction. 
Our simulation studies indicate that the stable model estimates ancestral states with reduced error in comparison to the homogenous Ornstein-Uhlenbeck model, when traits evolve by accumulating increments from a probability distribution without fixed variance (Table \ref{OUError}). The homogenous Ornstein-Uhlenbeck reconstruction of mammalian body mass (Table 4) is in some respects intermediate between the Brownian motion and stable reconstructions, with relatively small ancestors of Orders with small body size and relatively large ancestors of Orders with large body size. This can be interpreted in terms of the stabilizing selection model with an intermediate phylogenetic mean: the proposal of a small ancestral rodent (for example) permits a tendency to evolve back toward the mean within the rodent clade, while the proposal of a large ancestral carnivore (for example) permits the same tendency in the opposite direction within the carnivore clade. This pattern is most striking in the 
ancestral state inferred for Afrotheria (8.95kg versus 1.73kg under Brownian motion and 0.36kg under the stable model), where a large ancestor reduces the rate of evolution on branches leading ultimately to the large elephants and manatees while permitting many high-likelihood reductions of size toward the phylogenetic mean in 
small taxa such as Afrosoricida and Macroscelidea. This reconstruction for Afrotheria seems unlikely and may lead us to suppose that a single phylogenetic mean for the entire Eutheria does not form a realistic model of stabilizing selection. 

\citet{CooperPurvis2010} report success in fitting more complex heterogenous models to a larger set of mammalian body mass data. The Ornstein-Uhlenbeck model is easily extended to the heterogenous case by permitting more than one clade-specific phylogenetic  means \citep{Hansen1997,ButlerKing04,Hansenetal2008,Beaulieuetal2012}, the number and phylogenetic position of such means being specified \textit{a priori} or estimated from the data. Various transformations of the phylogeny, such as raising all branch lengths to a constant power in order to approximate speciational change \citep{Pagel1999} or somewhat \textit{ad hoc} transformation of branch lengths to maximize the likelihood of a Brownian model \citep{Grafen89, GittlemanKot1990, Garland92} have also been used to generate implicitly heterogenous models. The Early Burst model \citep{Blomberg2003,Harmonetal2010}, also applied by \citet{CooperPurvis2010}, is interesting in generating rate heterogeneity by allowing the rate of a Brownian motion to 
process to vary over time, rather than across branches or clades. The rate of evolution is taken to be an exponentially increasing or decreasing function of node height, allowing a greater proportion of evolutionary change to occur early in the phylogeny or late in the phylogeny depending on the choice of an exponential scaling factor $r$. We applied the Early Burst model to our mammalian body mass dataset but did not identify a significant deviation from $r=0$; ancestral state estimates in Table 4 are derived from a reconstruction based on $r=-0.009$ estimated by \citet{CooperPurvis2010}. The concentration of more evolutionary change in basal branches of the phylogeny appears to allow rapid early deviation from the phylogenetic mean value, with the majority of ancestral taxa exhibiting marginally smaller body sizes, often more consistent with the fossil evidence, than under the Brownian motion reconstruction, yet surprisingly with considerably less ordinal-level diversification than 
imputed by the stable model which does not build early diversification in to the stochastic process itself.

Homogenous approaches offer a number of advantages over heterogenous approaches in that the latter category must not only infer the parameters of the stochastic process but also must infer the structure of rate heterogeneity over the phylogeny. Especially when heterogenity is associated with clades, overfitting heterogenous models by positing too many rate shifts or clade-specific evolutionary regimes may become a danger. \citet{Eastmanetal2011} have proposed a heterogenous model of trait evolution that explicitly avoids overfitting by sampling over model parameter values and the number of model parameters simultaneously. The model is an extension of the standard Brownian motion model in which the rate of evolution is ``inherited'' over time but may undergo occasional shifts in value. Each shift introduces a new parameter that is penalized in a reversible jump Markov chain Monte Carlo algorithm. We found that the complexity of the reversible jump algorithm considerably increases the computational burden of 
fitting the model: for the body mass data considered here, the stable slice sampler accomplished around 70,000 steps per minute on a modest dual-core home laptop, versus around 2,500 per minute for the multi-rate model, without the need for a lengthy calibration of proposal densities beforehand. Ancestral states reconstructed under the Eastman model (Table 4) were in broad agreement with other non-Brownian methods presented here in imputing smaller early mammals, and were in close agreement with results of the Early Burst analysis.

In general, the stable model suggests a greater degree of ordinal-level diversification of mammalian body masses, and appears to accommodate a more volatile evolutionary process, than any of the other models considered here. For 13 of the 22 nodes listed in Table 4 the stable model reconstructs the smallest body masses of any model, and for 4 nodes it reconstructs the largest body mass of any model, making the stable reconstruction consistent with the hypothesis of small early mammals and occasional marked ordinal-level enlargement. The ability of the stable model to accommodate striking variation in evolutionary rate, even more so than approaches such as that of \citet{Eastmanetal2011} explicitly designed to model such variation, is most apparent in the highly diverse and species-poor Afrotheria, where the stable reconstruction involves the largest ancestral elephants and manatees yet the smallest Afroinsectivores of any of the methods considered here. While the heterogenous Ornstein-Uhlenbeck model binds 
rate volatility to the structure of the phylogeny through the assumption of clade-specific phylogenetic means, and the Eastman et al. RJ-MCMC model binds rate volatility to the structure of the phylogeny through the ``inheritance'' of rate shifts from ancestral to descendant branches, the stable model through its homogenously heavy tails provides unstructured volatility that is able to concentrate the production of evolutionary variation onto relatively few branches scattered across the phylogeny. 

The stable model is the simplest non-Brownian model considered here, requiring only a single parameter in addition to the standard Brownian motion model. The relative efficiency of the estimation procedure used to fit the stable model may make it attractive for analysis of very large trees or large sets of trees derived from Bayesian phylogenetics. Furthermore, deviation 
from the Brownian model according to the BPIC criterion may be used to provide independent statistical support for the adoption of one of the more complex heterogenous models currently available. Rates imputed by the stable model may guide appropriate selection of branches for independent rates in such cases. In the mammal body mass data examined here, for example, the frequency distribution of standardized trait changes along branches of the phylogeny (reported by the accompanying software) indicates accelerated evolution at the origin of a number of clades including \textit{Hyomys} (white-eared giant rats), Tragulidae (mouse deer), Manidae (pangolins), Megachiroptera (megabats), Megadermatidae (false vampire bats), Solenodontidae (solenodons), Orycteropodidae (aardvark) and Hyracoidea (hyraxes), suggesting that these clades may merit their own phylogenetic mean values under a heterogenous Ornstein-Uhlenbeck approach. In order to determine whether such a model is useful in any particular case it is 
necessary, as with any stochastic model of evolution, to rigorously constrain the model empirically, and while the results presented here are primarily illustrative and to provide comparison across unconstrained models, we note that the low ancestral state inferences for extinct taxa at the root of Rodentia, Lagomorpha, Primates, Chiroptera and Lipotyphla, and the high ancestral state inferences for taxa at the root of Sirenia and Elephantidae, appear broadly in line with fossil evidence.

While our homogenous approach may be associated with some advantages with respect to efficiency, simplicity and unstructured volatility, the heterogenous models such as Early Burst have the benefit of imposing an explicit evolutionary narrative on the process of trait diversification which may be useful for exploring and testing general hypotheses about historical processes \citep{Harmonetal2010}. Heterogenous models typically involve the elaboration of a simple Gaussian kernel to accommodate phylogeny- or time-structured variation in the evolutionary process. We suggest that in future work heterogenous stable models analogous to those considered above may be readily generated by directly replacing the Gaussian kernel with the more general stable kernel, at the expense of a single parameter. Stable Ornstein-Uhlenbeck processes, for example, are already well-characterized \citep{SamorodnitskyTaqqu94}. The stable model we introduce to phylogenetic evolutionary biology here may find other uses, for example in 
assigning substitution rates to edges on phylogenetic trees under relaxed 
clock models \citep{ThorneKishinoPainter98}. One primary obstacle to the replacement of Gaussian kernels by stable kernels in models of continuous character evolution is that that stable distributions have undefined variance \citep{SamorodnitskyTaqqu94}. Methods making direct use of variance are typically used to detect correlated evolution between multiple continuous characters evolving on the same phylogenetic tree \citep{Martins96}. Independent contrasts \citep{Felsenstein85} for example, generates standardized data points for each univariate character by scaling the increments accruing along paired branches of a phylogeny by the square root of the sum of branch lengths, which is proportional to the expected standard deviation of a Brownian process. Methods of phylogenetic regression \citep{Grafen89,Garlandetal93,MartinsHansen97} extend least squares methods to multivariate phylogenetic data by incorporating branch length and topological information into the model's covariance matrix. The fact that stable 
variance is undefined means that there is no stable equivalent to standard 
deviation or the covariance matrix. We note that regression and correlation models based on stochastic processes driven by non-Gaussian stable perturbations have been implemented successfully in non-phylogenetic fields \citep[i.e., ][]{McCulloch98b,Frain08,Paulaauskas03}. These approaches raise the prospect that likelihood-based analysis of heavy tailed multivariate distributions may offer useful insights into future studies of correlated evolution of multiple continuous characters in evolutionary biology, since correlated evolution is precisely the kind of problem domain in which the putative Brownian assumptions of neutrality and gradualism are likely to be invalid.

\section*{Acknowledgments}
The authors thank Gavin Thomas for providing invaluable criticism of the model and performance of the software implementation for large datasets, Tomohiro Ando for providing material on BPIC, and members of the UBC and SFU evolution groups for their advice on an early version of this manuscript. This work was enabled by computational resources provided by Compute/Calcul Canada and the kind support of Brian Corrie at IRMACS. ME was financially supported by NSERC and the Human Evolutionary Studies Program at SFU. 

\bibliographystyle{sysbio}
\bibliography{biblio}{}

\clearpage

\begin{table}
\begin{center}
\begin{tabular}{rc|cccccccccc}
&& \multicolumn{9}{|c}{Tree size (tips)} \\ 
Error&\multicolumn{1}{c|}{$\alpha$} & 25 & 40 & 60 & 90 & 130 & 175 & 235 & 325 & 440 & 600 \\ \hline
& \multicolumn{1}{c|}{}& & & \\
Type I&\multicolumn{1}{c|}{2.0}                    & 0.11 & 0.08 & 0.04 & 0.01 & 0.01 & 0.00 & 0.00 & 0.00 & 0.00 & 0.00 \\[0.2cm].
\multirow{5}{*}{Type II} &\multicolumn{1}{c|}{1.8} & 0.66 & 0.45 & 0.47 & 0.51 & 0.24 & 0.24 & 0.20 & 0.10 & 0.07 & 0.04 \\
&\multicolumn{1}{c|}{1.6}                          & 0.39 & 0.20 & 0.08 & 0.21 & 0.04 & 0.01 & 0.00 & 0.00 & 0.00 & 0.00 \\
&\multicolumn{1}{c|}{1.4}                          & 0.16 & 0.04 & 0.00 & 0.10 & 0.00 & 0.00 & 0.00 & 0.00 & 0.00 & 0.00 \\
&\multicolumn{1}{c|}{1.2}                          & 0.09 & 0.00 & 0.00 & 0.04 & 0.00 & 0.00 & 0.00 & 0.00 & 0.00 & 0.00 \\
&\multicolumn{1}{c|}{1.0}                          & 0.02 & 0.00 & 0.00 & 0.00 & 0.00 & 0.00 & 0.00 & 0.00 & 0.00 & 0.00 \\ 
\end{tabular}
\end{center}
\caption{Type I and Type II error rates resulting from stable versus Brownian model selection using the BPIC model selection criterion, with data simulated on trees of various sizes under stable processes with varying indices of stability.}
\label{BPIC}
\end{table} 

\clearpage

\begin{table}
\begin{center}
\begin{tabular}{c|cccccccccc}
& \multicolumn{10}{c}{Tree size (tips)} \\ 
$\alpha$ & 25 & 40 & 60 & 90 & 130 & 175 & 235 & 325 & 440 & 600
\\[0.5ex] \hline\\[-1.5ex]
2.0 & 1.00 & 1.02 & 1.01 & 1.00 & 1.01 & 1.00 & 1.00 & 1.00 & 1.00 & 1.00 \\
1.8 & 0.97 & 0.96 & 0.94 & 0.92 & 0.84 & 0.82 & 0.81 & 0.77 & 0.76 & 0.73 \\
1.6 & 0.85 & 0.79 & 0.70 & 0.64 & 0.61 & 0.56 & 0.47 & 0.47 & 0.42 & 0.37 \\
1.4 & 0.66 & 0.50 & 0.49 & 0.35 & 0.28 & 0.29 & 0.20 & 0.24 & 0.16 & 0.16 \\
1.2 & 0.48 & 0.36 & 0.26 & 0.15 & 0.14 & 0.12 & 0.11 & 0.09 & 0.08 & 0.07 \\
1.0 & 0.21 & 0.16 & 0.08 & 0.07 & 0.06 & 0.04 & 0.03 & 0.02 & 0.02 & 0.01 \\
\end{tabular}
\end{center}
\caption{Sum of squared errors in ancestral state reconstruction, median ratio of stable/Brownian reconstruction error, with data simulated on trees of various sizes under stable processes with varying indices of stability.}
\label{BrownianError}
\end{table}

\clearpage

\begin{table}
\begin{center}
\begin{tabular}{c|cccccccccc}
& \multicolumn{10}{c}{Tree size (tips)} \\ 
$\alpha$ & 25 & 40 & 60 & 90 & 130 & 175 & 235 & 325 & 440 & 600 
\\[0.5ex] \hline\\[-1.5ex]
2.0 & 0.36 & 0.35 & 0.35 & 0.33 & 0.43 & 0.52 & 0.54 & 0.71 & 0.75 & 0.83 \\
1.8 & 0.34 & 0.33 & 0.36 & 0.36 & 0.38 & 0.44 & 0.50 & 0.48 & 0.56 & 0.47 \\
1.6 & 0.31 & 0.28 & 0.29 & 0.31 & 0.31 & 0.46 & 0.27 & 0.24 & 0.27 & 0.24 \\
1.4 & 0.28 & 0.24 & 0.25 & 0.18 & 0.16 & 0.17 & 0.13 & 0.14 & 0.11 & 0.11 \\
1.2 & 0.23 & 0.17 & 0.16 & 0.11 & 0.09 & 0.08 & 0.07 & 0.06 & 0.06 & 0.04 \\
1.0 & 0.17 & 0.11 & 0.08 & 0.06 & 0.04 & 0.03 & 0.03 & 0.02 & 0.02 & 0.01 \\

\end{tabular}
\end{center}
\caption{Sum of squared errors in ancestral state reconstruction, median ratio of stable/Ornstein-Uhlenbeck reconstruction error, with data simulated on trees of various sizes under stable processes with varying indices of stability.}
\label{OUError}
\end{table}

\clearpage
\newgeometry{left=1cm,right=1cm,bottom=3cm,top=3cm}
\begin{landscape}
\begin{table}
\begin{center}

\begin{tabular}{lcccccl}
Clade&\multicolumn{5}{c}{Reconstructed body mass (kg)} & Fossil estimates (kg) \\
&BM&OU&EB&Eastman&Stable& \\[0.2cm]
\cline{1-7}
& \multicolumn{1}{c}{}&&&&&\\

Eutheria & 1.24 & 1.23 & 1.14  & 0.97 & 0.33 & \parbox{5cm}{0.015-0.017 (\textit{Juramaia}) \\0.02-0.025 (\textit{Eomaia})\\Shrew-sized (\textit{Prokennalestes})} \\

\hspace{0.5cm}Euarchontoglires & 0.72 & 0.43 & 0.73 & 0.60 & 0.34 & \\
\hspace{1.0cm}Rodentia & 0.37 & 0.26 & 0.39  & 0.34 & 0.24 & \parbox{5cm}{0.0003 (\textit{Tribosphenomys}) \\ 0.0014 (\textit{Microparamys}) }\\
\hspace{1.0cm}Lagomorpha & 0.53 & 0.36 & 0.56 & 0.50 & 0.37& 0.1-0.2 (Lagomorpha indet.) \\
\hspace{1.0cm}Primates & 0.76 & 0.49 & 0.77 & 0.64 & 0.45 & 0.1 (\textit{Purgatorius}) \\
\hspace{1.0cm}Scandentia & 0.25 & 0.15 & 0.27 & 0.25 & 0.25 & \\

\hspace{0.5cm}Laurasiatheria & 1.13 & 0.61 & 1.10 & 0.60 & 0.25 & \\
\hspace{1.0cm}Cetartiodactyla & 23.6 & 80.29 & 17.60 & 18.34 & 79.22 & \parbox{5cm}{2.2-3.9 (Diacodexeidae) \\2.6-3.9 (Homacodontidae)\\6.6-9.6 (\textit{Laredochoerus})}\\
\hspace{1.0cm}Perissodactyla & 86.5 & 152.51 & 57.17 & 72.75 & 273.76 & \parbox{5cm}{158.2 (\textit{Hyrachyus}) \\ 740.1 (\textit{Amynodon})}\\
\hspace{1.0cm}Carnivora & 6.34 & 8.77 & 5.45 & 5.23 & 15.20 & \parbox{5cm}{0.15-10 (Viveravidae) \\ 1-10 (Miacidae)} \\ \\
\hspace{1.0cm}Pholidota & 4.61 & 5.54 & 4.48 & 4.47 & 4.28 & \\
\hspace{1.0cm}Chiroptera & 0.08 & 0.12 & 0.09 & 0.02 & 0.02 & 0.012-0.015 (\textit{Icaronycteris})\\
\hspace{1.0cm}Eulipotyphla & 0.58 & 0.41 & 0.62 & 0.36 & 0.22 & 0.011 (\textit{Paranyctoides})\\

\hspace{0.5cm}Xenarthra & 3.04 & 3.13 & 2.60 & 2.8 & 2.15 & \\
\hspace{1.0cm}Pilosa & 4.17 & 3.94 & 3.87 & 2.80 & 4.49 & \\
\hspace{1.0cm}Cingulata & 3.30 & 3.36 & 3.12 & 3.23 & 2.40 & \\

\hspace{0.5cm}Afrotheria & 1.73 & 8.95 & 1.50 & 1.51 & 0.36 & \\
\hspace{1.0cm}Afrosoricida & 0.72 & 0.17 & 0.77 & 0.65 & 0.15 & \\
\hspace{1.0cm}Macroscelidea & 0.30 & 0.44 &0.34 & 0.65 & 0.17 & \\
\hspace{1.0cm}Hyracoidea & 3.40 & 9.14 & 3.38 & 3.38 & 3.56 & $>4$ (\textit{Heterohyrax}) \\
\hspace{1.0cm}Proboscidea & 1,478.15 & 949.76 & 1039.24 & 1,460.62 & 2,928.50 & $>2,000$ (\textit{Palaeomastodon}) \\
\hspace{1.0cm}Sirenia &90.33 & 204.84 & 53.41 & 87.24 & 341.12 & 488.1 (\textit{Halitherium}) \\
\end{tabular}

\caption{Ancestral state reconstruction of adult female body mass (kg) based on a dataset of 1,679 eutherian mammal species means, under Brownian Motion (BM; maximum likelihood), Ornstein-Uhlenbeck (OU; maximum likelihood), Early Burst (EB; maximum likelihood), Eastman \textit{et al}'s (2011) heterogenous multi-rate (maximum posterior probability) and stable (maximum posterior probability) models. Reconstructed values are provided for the most recent common ancestors of extant taxa in the specified clades. The right column details a selection of oldest fossil taxa within each clade for which body mass estimates are available (data from \citet{PBDB,Fleagle99,GoswamiFriscia2010,Ji02,Luoetal2011,Rasmussen96,Rose06,Rose08,RydellSpeakman95,Wood10})}
\end{center}
\label{MammalAncStates}
\end{table}
\end{landscape}
\restoregeometry

\clearpage

\begin{figure}[p]
\centering
\includegraphics[scale=0.4]{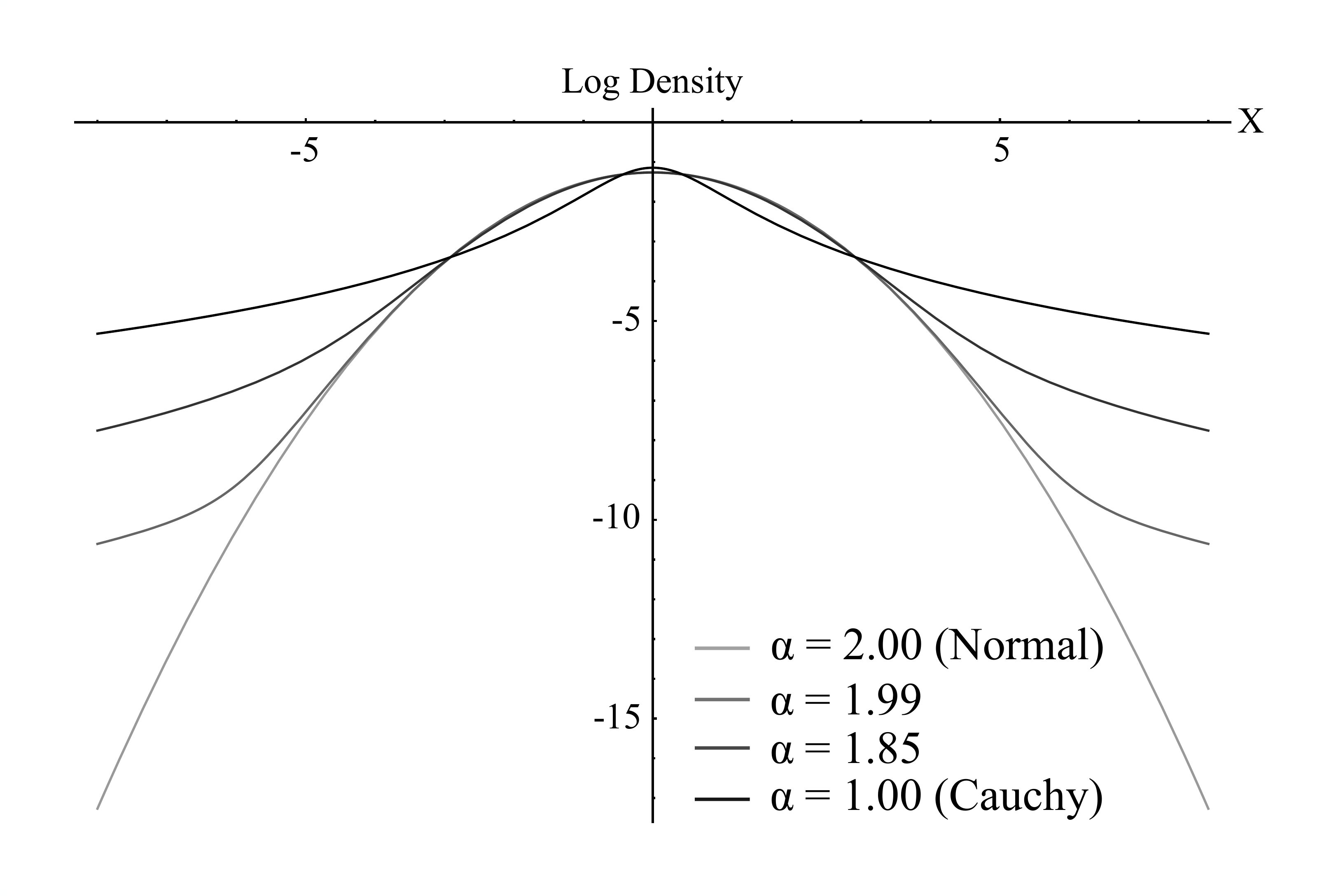}
\caption{Log probability densities of stable distributions with varying $\alpha$, including the normal distribution ($\alpha$=2) and the Cauchy distribution ($\alpha=1$) as special cases.}
\end{figure}

\clearpage

\begin{figure}[p]
\centering
\includegraphics[scale=0.4]{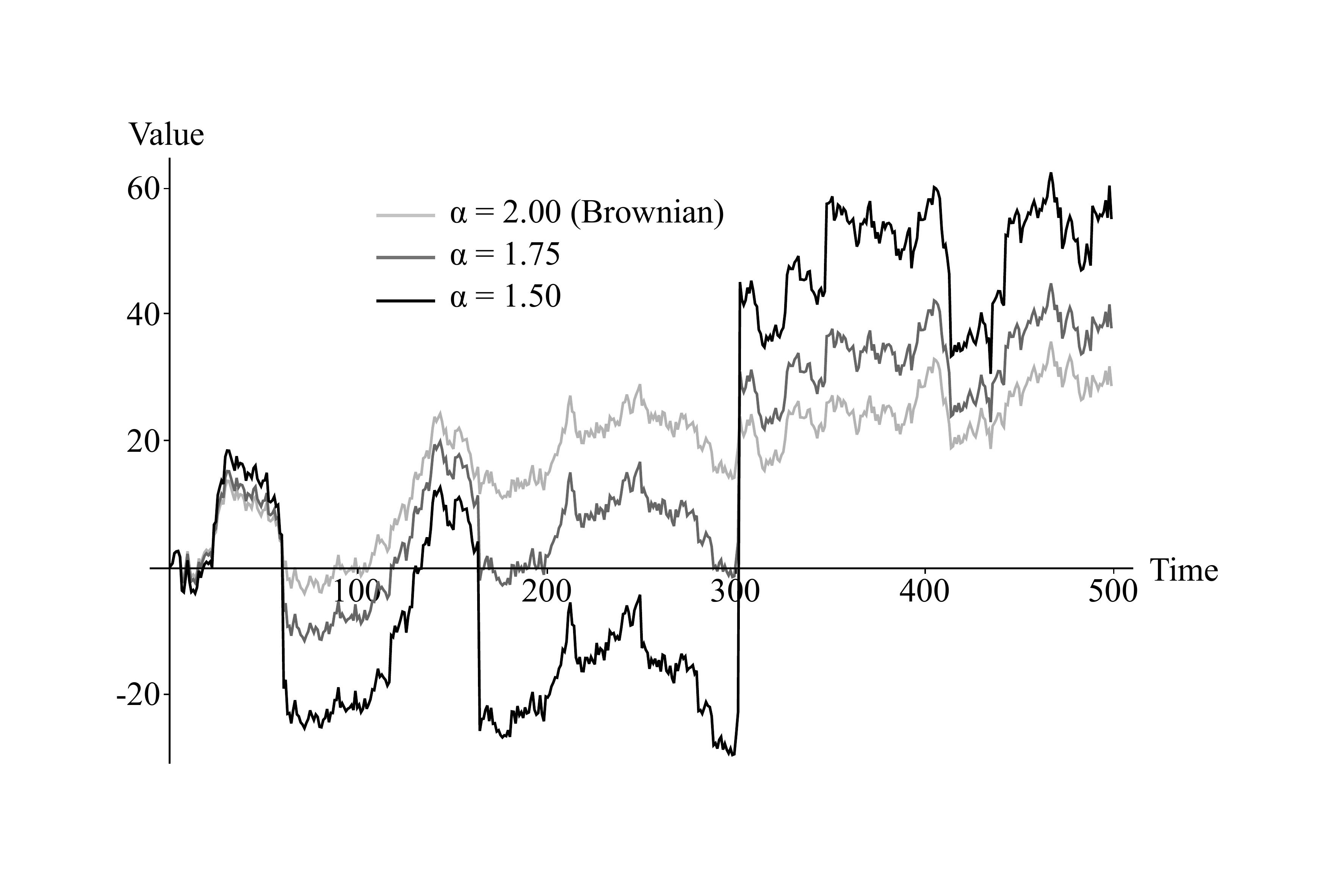}
\caption{Stable random walks with varying $\alpha$, driven by accumulation of a single sample of 500 increments drawn from a uniform distribution between zero and one, transformed onto stable distributions using the inversion method \citep{Devroye86}.}
\end{figure}
 
\clearpage

\begin{figure}[p]
\centering
\includegraphics[scale=0.3]{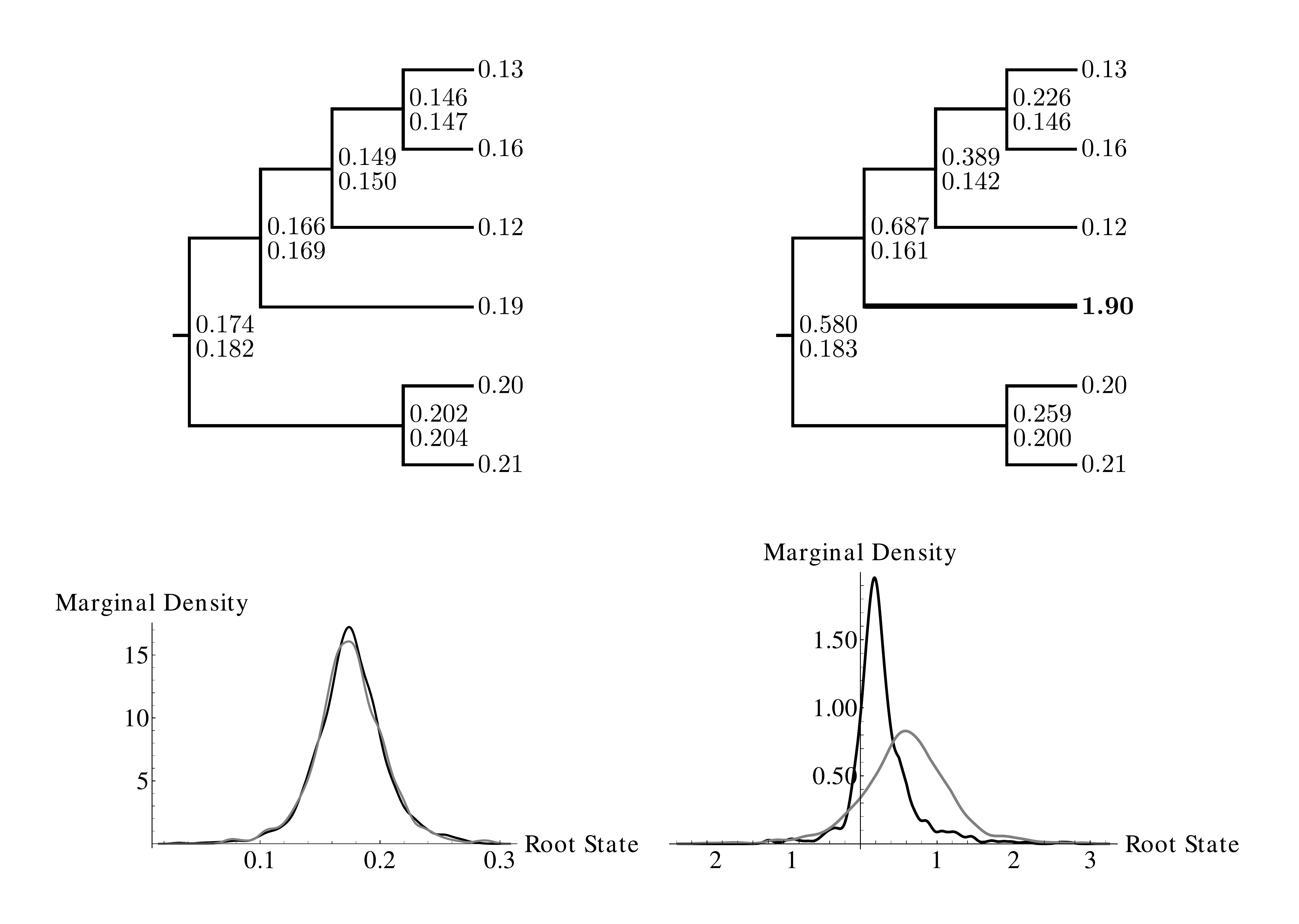}
\caption{Ancestral state reconstruction of data simulated under Brownian motion. Top row: phylogenies exhibiting ancestral state reconstructions; each tip is labelled with the known character state, while at each internal node, upper values represent the Brownian motion maximum likelihood reconstruction and lower values represent the reconstruction under the stable Markov chain Monte Carlo model to be described in this paper (left: original simulated data; right: the character value of a single tip has been multiplied by ten to simulate rapid evolution (or measurement error) on a single branch). Bottom row: the marginal probability density derived from MCMC for the ancestral state assignment of the root node using unmodified (left) and modified (right) data; the Brownian motion marginal probability density is grey and the stable margin probability density is black.}
\end{figure}

\clearpage

\begin{figure}[p]
\centering
\includegraphics[scale=0.3]{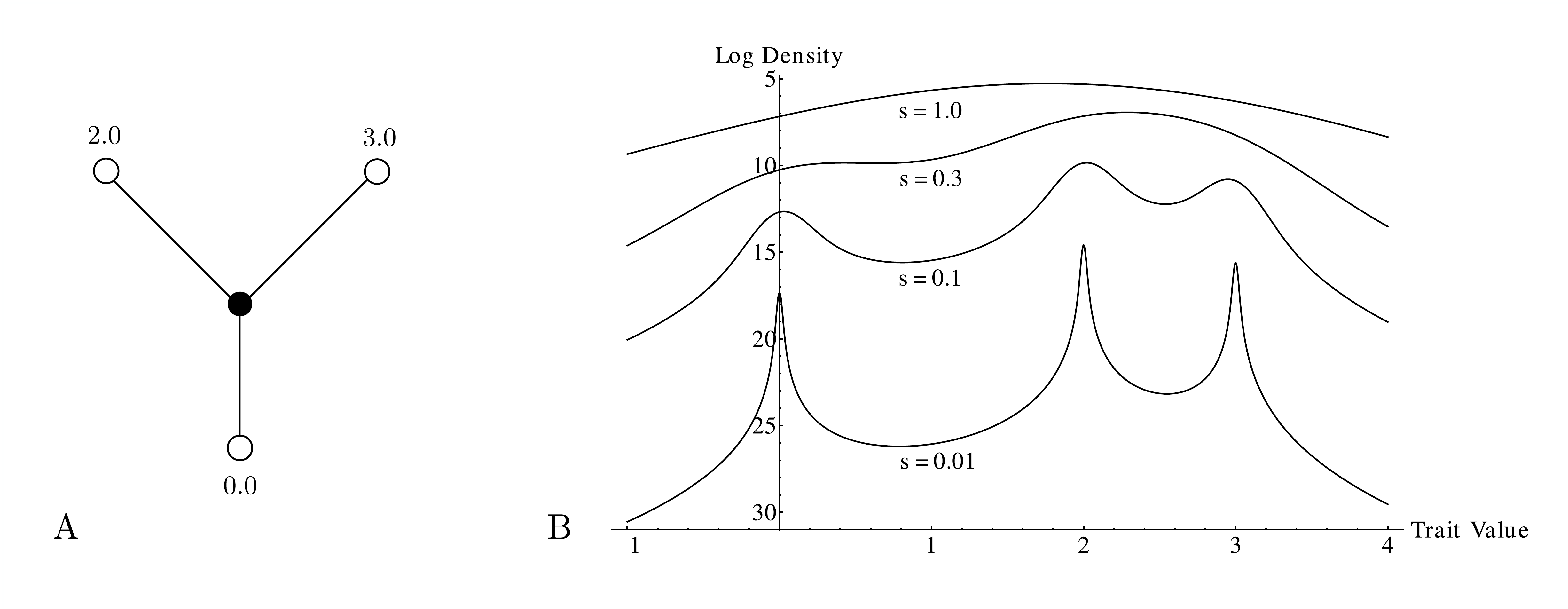}
\caption{(A) A focal node in a phylogenetic tree has two children with trait values 2 and 3, and a parent with trait value 0, to which it is connected by branches of unit length. (B) The potentially multimodal conditional probability density function describing the probability with which the focal node has some trait value given the values of its children and parent (several curves differ in terms of the scale parameter of the stable distribution, but have a common $\alpha = 1.5$).}
\end{figure}

\clearpage

\begin{figure}[p]
\centering
\includegraphics[scale=0.4]{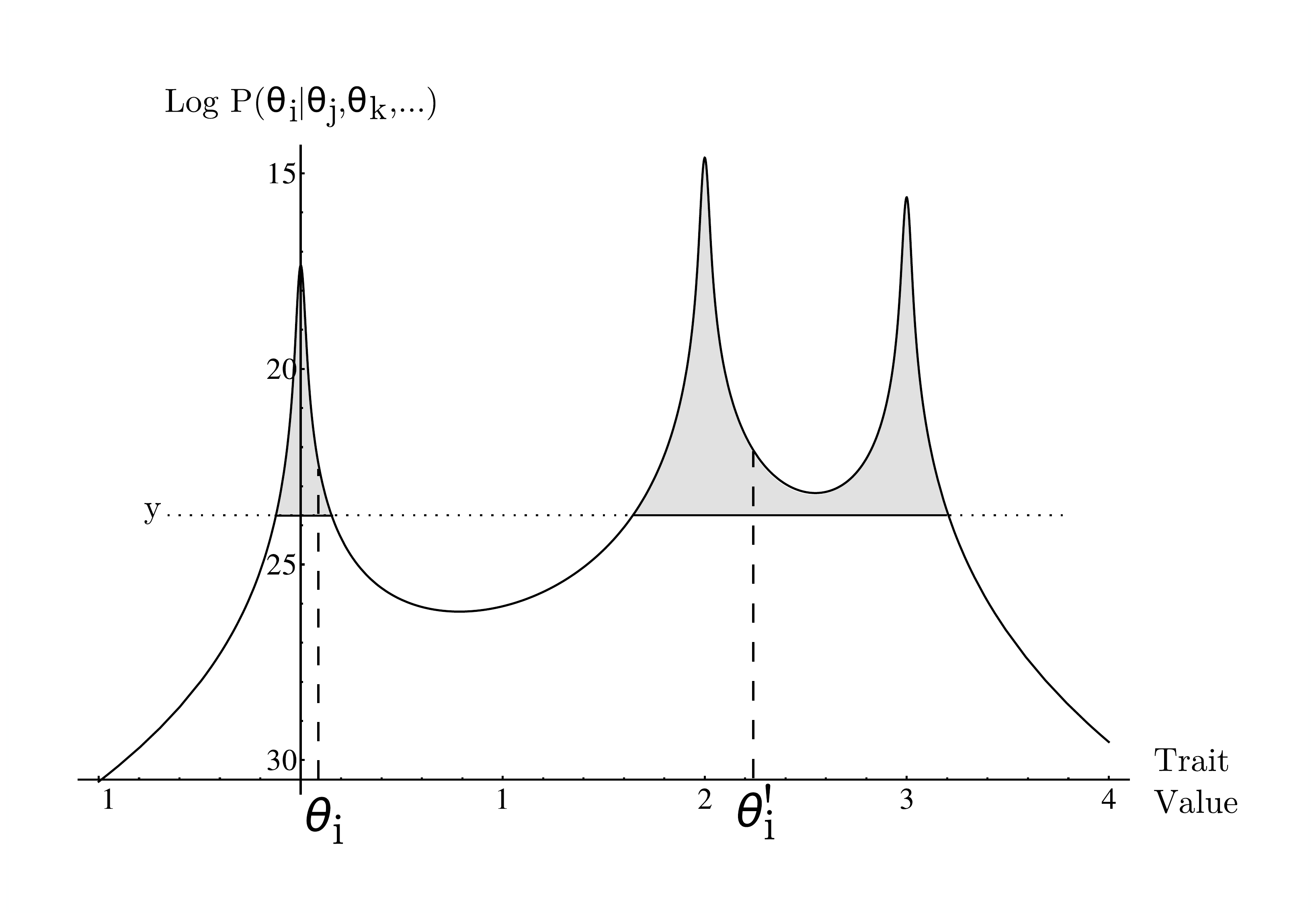}
\caption{An example of slice sampling updating the value of the focal node from Figure 4(A) from $\theta_i$ to $\theta_i^\prime$. First, the conditional probability that the value of the evolving character at the focal node is equal to $\theta_i$ --- given the other ancestral states and stable model parameters --- is calculated. A random number $y$ is drawn from the uniform distribution between zero and this conditional probability (marked with a dotted line). The set of possible values of $\theta_i^\prime$ is estimated by bracketing regions around the modes of the distribution for which the conditional probability is greater than $y$ (solid lines and shaded region of the distribution). The new value $\theta_i^\prime$ is drawn as a uniform random variable from this set.}
\end{figure}

\end{document}